\documentclass[aps,prl,twocolumn,showpacs,preprintnumbers,amsmath,amssymb,
floatfix,superscriptaddress]{revtex4-1}

\usepackage{ulem}

\usepackage{graphics}
\usepackage{graphicx}
\usepackage{amsmath}
\usepackage{setspace}
\usepackage{braket}
\usepackage{epstopdf}
\usepackage{comment}
\usepackage{hyperref}
\usepackage{bm}
\usepackage{xfrac}
\usepackage{blindtext}
\usepackage{mathrsfs}
\usepackage{xcolor}
\usepackage[version=4]{mhchem}

\newcommand{\affA}{Van der Waals-Zeeman Institute, Institute of Physics, University of Amsterdam, 1098 XH Amsterdam, Netherlands}
\newcommand{\affB}{QuSoft, Science Park 123, 1098 XG Amsterdam, the Netherlands}
\newcommand{\affC}{Institute for Theoretical Physics, Institute of Physics, University of Amsterdam, Science Park 904, 1098 XH Amsterdam, the Netherlands}

\newcommand{\affF}{Department of Physics, Stony Brook University, Stony Brook, New York 11794, USA}
\newcommand{\affG}{Institute for Advanced Computational Science, Stony Brook University, Stony Brook, New York 11794, USA}

\begin{document}

\title{Trap-assisted complexes in cold atom-ion collisions}

\author{H.~Hirzler}\affiliation{\affA}
\author{E.~Trimby}\affiliation{\affA}
\author{R.~Gerritsma}\affiliation{\affA}\affiliation{\affB}
\author{A.~Safavi-Naini}\affiliation{\affB}\affiliation{\affC}
\author{J.~P\'erez-R\'{i}os}\affiliation{\affF}\affiliation{\affG}

\date{\today}

\begin{abstract}
We theoretically investigate the trap-assisted formation of complexes in atom-ion collisions and their impact on the stability of the trapped ion. The time-dependent potential of the Paul trap facilitates the formation of temporary complexes by reducing the energy of the atom, which gets temporarily stuck in the atom-ion potential. As a result, those complexes significantly impact termolecular reactions leading to molecular ion formation via three-body recombination. We find that complex formation is more pronounced in systems with heavy atoms, but the mass has no influence on the lifetime of the transient state. Instead, the complex formation rate strongly depends on the amplitude of the ion's micromotion. We also show that complex formation persists even in the case of a time-independent harmonic trap. In this case, we find higher formation rates and longer lifetimes than the Paul trap, indicating that the atom-ion complex plays an essential role in atom-ion mixtures in optical traps. 
\end{abstract}

\maketitle

\paragraph{Introduction}

Most chemical reactions occur via the formation of an intermediate complex that facilitates the reagents to transform into products. These intermediate complexes can be viewed as quasi-bound states of the reagents that, via internal energy exchange, may evolve into the products of the reaction. However, those complexes relevant for atmospheric chemical kinetics~\cite{Troe1994} or biomolecular reactions ~\cite{Su2013, Bjork2016, Hu2019}, typically show a lifetime $\lesssim$1~ns, which makes it very hard to observe them directly. On the contrary, in the ultracold regime, it has been shown that bi-molecular reactions show long-lived complexes that can be observed and diagnosed~\cite{observation,observation_bis,observationHamburg,observationRbCs}. 

Hybrid ion-atom systems present a perfect arena to study collisions between ions and neutral species~\cite{zipkes:2010b, Schmid:2010, Ravi:2012, haze:2015, Saito:2017, Joger:2017, Sikorsky:2018, Li2020, mahdian2021, Mohammadi2021, Hirzler2022, Katz2021}. In those systems, atom-ion complexes have been predicted~\cite{Cetina:2012, meir2017} due to the time-dependent trapping potential for the ion. However, a systematic study on the properties of atom-ion complexes is still lacking, as is their effect on reactive processes such as ion-atom-atom three-body recombination. Ion-atom-atom three-body recombination is a termolecular reaction in which three free atoms collide to form a molecule and a free atom as products. Such a reaction can be viewed as the result of two bimolecular processes: first, two particles collide to form a complex; second, a third particle collides with the complex and stabilizes it. This model's reaction rate depends on the competition between the complex's lifetime and the colliding partners' collision time. As a result, if it is possible to modify the lifetime of the complex, it will be plausible to control the ion-atom-atom three-body reactivity, thus, opening a new avenue for controlled chemistry in hybrid atom-ion systems without requiring reaching the ultracold regime for the atom-ion scattering.

This Letter presents a theoretical study on atom-ion complex formation in time-dependent and time-independent traps. We show that it is possible to control the lifetime of the complexes and, with it, three-body recombination reactions. In particular, we find that the complex formation probability depends on the atom mass but has a minor effect on the lifetime of the intermediate states. Additionally, we show that not only the formation of quasi-bound states persists in static harmonic traps, but that these complexes have higher formation probabilities and longer lifetimes than in the Paul trap case. Finally, in the Paul trap, we study the effect of the micromotion amplitude on the formation of the intermediate states. Our study covers a vast trap parameter space in atomic and ionic species, thus offering a roadmap to control ion-atom complexes' lifetimes and observation. 

%%%%%%%%%%%%%%%%%%%m intro %%%%%%%%%%%%%%%%%%%%%%%%%%%%%

\begin{figure*}
	\centering
	\includegraphics[width=\linewidth]{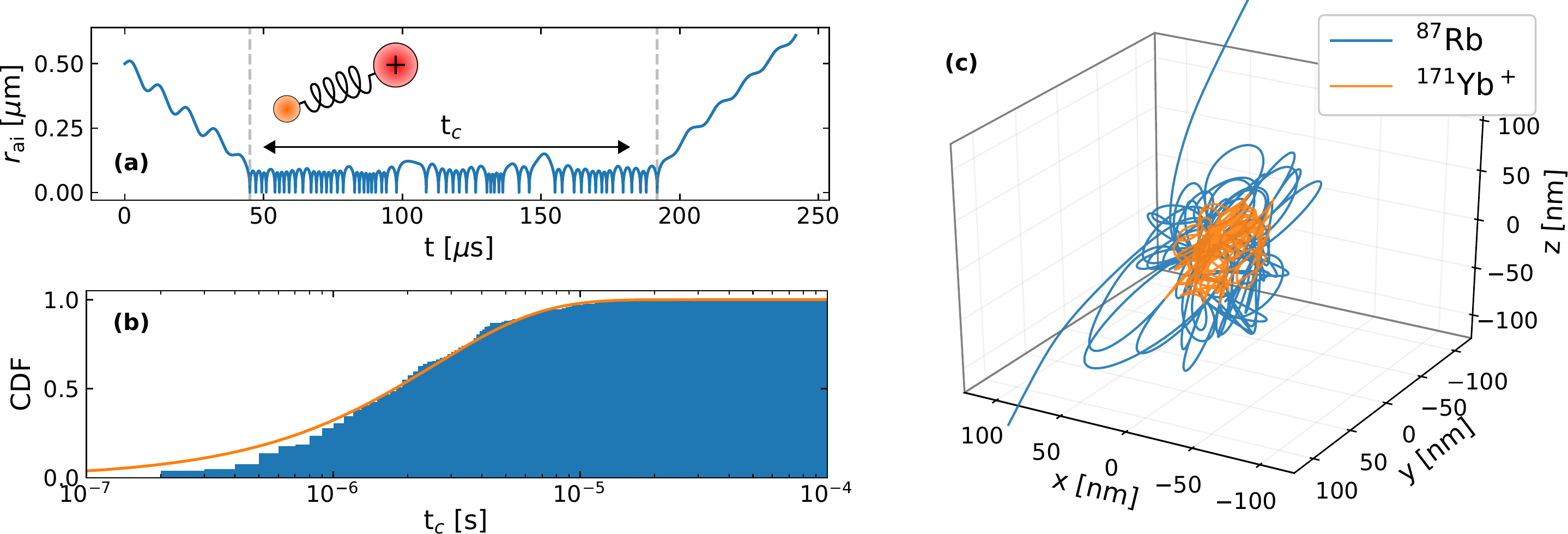}
 	\caption{Trajectory of a $^{87}$Rb atom colliding with a $^{171}$Yb$^+$ ion in a time-independent harmonic ion trap with trap frequency $\omega_\mathrm{3D}=100\,$kHz, $T_a=0.5\,\mu$K and $T_i=10\,\mu$K. Panel (a) shows the atom-ion distance as a function of time, showing the formation of a complex of lifetime t$_c=10.3\,\mu$s. Panel (b) shows a characteristic cumulative distribution function for the lifetime of the complex, wherein the solid line represents a fitting to an exponential function (see text for details). Panel (c) shows the same trajectory as in panel (a) in Cartesian coordinates.}
 	\label{fig1}
\end{figure*}

%%%%%%%%%%%%%%%%%%%%%%%%%%%%%%%%%%%%%%%%%%%%%%%%%
\paragraph{Theoretical approach}
To simulate the dynamics of atom-ion collisions in the presence of a trap, we use classical trajectory calculations. Typically, the atom-ion $s$-wave limit is orders of magnitude below the collision energy~\cite{Tomza:2019,BookJPR} and large numbers of partial waves contribute. Hence, classical approaches are justified. In addition, in the case of the Paul trap, the deep time-dependent electric trap has a strong impact on the collisions in it, which complicates approaches based on quantum mechanical methods.

In atom-ion systems, the charge-induced dipole interaction results in an attractive long-range $-\alpha/2r^4$ interaction, wherein $\alpha$ is the atom polarizability, and $r$ stands for the interparticle separation. Collisions in the presence of a trap are better described by the distance of the closest approach $b=\mathrm{min}(r(t))$. For $b>b_\mathrm{L}$, we find elastic collisions which allow small energy transfer, whereas for $b<b_\mathrm{L}$ Langevin, inelastic and reactive collisions occur. Here, $b_\mathrm{L} = (2\alpha/E_\mathrm{col})^{1/4}$ is the Langevin impact parameter with collision energy $E_\mathrm{col}$, defining a capture radius for particles to visit the short-range interaction region, leading to efficient energy and momentum transfer. An example of a trajectory is shown in Fig.~\ref{fig1}.

In this scenario, the Langevin collision rate, $\Gamma_\mathrm{L}=2\pi n_a \sqrt{\alpha/\mu}$, is a function of the atom density $n_a$ and the reduced mass $\mu$. The full model atom-ion potential reads
\begin{equation}
	V_\mathrm{ai} =  \frac{C_6}{r^6} -\frac{\alpha}{2r^4} ,
\end{equation}
where $C_6$ is the repulsion coefficient as a consequence of electronic exchange-repulsion interactions and ultimately nuclear repulsion. Herein, we consider two kind of traps. On the one hand, a time-independent harmonic trap (HT) representing an ion in an optical trap~\cite{schneider:2010} given by  
\begin{equation}
	V_\mathrm{HT}(\vec{r}) = \frac{1}{2}\sum_{j=1}^{3}  m \omega_\mathrm{j}^2 r_{\text{ion}_j}^2, 
\end{equation}
where $j\in\{x,y,z\}$ is the direction, $r_{\mathrm{ion}_j}$ is the ion position and $\omega_j$ the trap frequency. On the other hand, a Paul trap (PT), which uses time-dependent electric fields for the ion confinement. The potential of the Paul trap is given by 
\begin{equation}
	V_\mathrm{PT}(\vec{r}_\mathrm{ion}, t) = \frac{U_\mathrm{dc}}{2} \sum_{j=1}^{3} \alpha_j r_{\mathrm{ion}_j}^2 + \frac{U_\mathrm{rf}}{2} \cos{(\Omega t)} \sum_{j=1}^{3} \alpha'_j r_{\mathrm{ion}_j}^2,
\end{equation}
where $r_\mathrm{ion} = (0, 0, 0)^\mathrm{T}$ is the trap center, $U_\mathrm{dc}$ and $U_\mathrm{rf}$ are the curvatures of the electric dc and rf fields, respectively, and $\alpha_j$ and $\alpha_j'$ are geometry factors. Here, we use $-2\alpha_1 =-2\alpha_2=\alpha_3$ and $\alpha_1'=-\alpha_2'=1$, $\alpha'_3=0$. In the radial direction, the ion oscillates with a slow secular motion with frequency $\omega_\perp \approx\Omega q/\sqrt{8}$ which is superimposed by a fast micromotion which oscillates at $\Omega$~\cite{Leibfried:2003}.

Every simulation initializes the atom distance $r_\mathrm{start}$ from the ion and randomizes the velocity vectors from thermal distributions~\cite{Fuerst:2018, Hirzler2020, Trimby2022}. During the collision, we identify the presence of a complex characterizing the classical inner turning point~\footnote{At the inner turning point, the forces maximize and the step-size reaches its lowest values, which reaches below an identifier value.}. Then, the complex lifetime t$_\text{c}$ is obtained by tracking the time between the first and last visit of the inner turning point, as shown in panel (a) of Fig.~\ref{fig1} for a $^{87}$Rb atom colliding with a $^{171}$Yb$^+$ ion. Finally, simulations stop after complex dissociation when the atom leaves the interaction sphere. From the numerical simulations we calculate the complex formation probability as
\begin{equation}
\label{eq4}
	P_\text{c} = \frac{N_\text{c}}{N_\text{L}}\pm\delta_{P_\text{c}}, \quad \delta_{P_\text{c}} = \sqrt{\frac{N_\text{c} \left(N_\text{L}-N_\text{c}\right)}{N_\text{L}^3}},
\end{equation}
where $N_\text{c}$ and $N_\text{L}$ are the number of events that result in a complex and in Langevin collisions, respectively. 

\begin{figure}
	\centering
	\includegraphics[width=\linewidth]{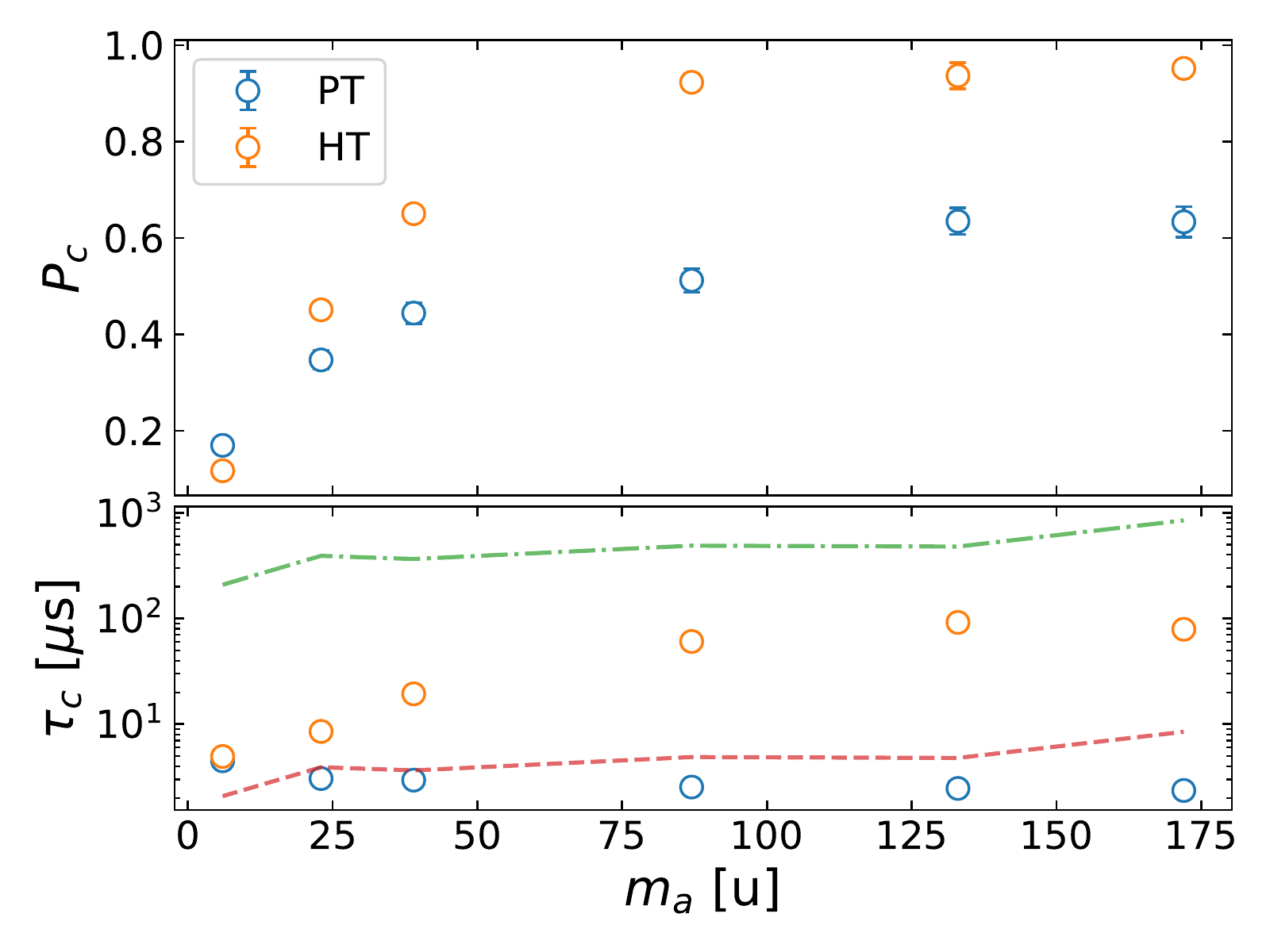}
 	\caption{Quasi-bound state formation probability (upper panel), and lifetime (lower panel), for collisions of (Li, Na, K, Rb, Cs, Yb) atoms and Yb$^+$ ions. The collision energy is $\approx15\,\mu$K ($T_\text{a}=10\,\mu$K, $T_\text{i}=10\,\mu$K). The simulations are done in a Paul trap (PT) and a harmonic trap (HT) with the parameters described in the main text. The error bars are estimated via Eq.(\ref{eq4}). Each data point corresponds to at least $10^5$ trajectories. The dashed-green and dashed-red curves correspond to $1/\Gamma_L$ for $n_a=$10$^{18}$~m$^{-3}$ and 10$^{20}$~m$^{-3}$, respectively. 
 	}
 	\label{fig:masses}
\end{figure}

\paragraph{Results}

We study atom-ion complex formation in a Paul trap (PT) and a harmonic trap (HT), finding that the formation rate increases towards heavier atoms, as shown in Fig.~\ref{fig:masses}. For these simulations, we use the Yb$^++X$ system with $X\in\{$Li,Na,K,Rb,Cs,Yb$\}$, and we adjust the mass and $\alpha$, while keeping the collision energy constant. Even for large mass ratios (Yb$^+$/Li), we find complex formation probabilities of $\approx15\,$\%. Remarkably, results for HT show a more significant complex formation probability than the PT for the whole set of parameters, even reaching almost 100$\%$ of probability for the heavier species. That is, every collision leads to an atom-ion complex.

%probability for complex formation approaches unity for heavier atoms. 
On the other hand, we explore the complex lifetime, $\tau_\text{c}$, which is shown in the lower panel of Fig.~\ref{fig:masses}. $\tau_{\text{c}}$ is calculated from the cumulative distribution function of events with complex lifetimes t$_c$ and then extracts the 1/$e$ value of a fitted exponential function, as it is shown in Fig.~\ref{fig1}. In the PT case, we find $\tau_{\text{c}}\sim 2\,\mu$s independently of the atom's mass. On the contrary, for the HT case, the atom's mass drastically impacts $\tau_{\text{c}}$ showing a wide range of values between 5 and 100$\mu$s. Next, we compare the complex lifetime versus the collisional time associated with Langevin collisions, i.e., the typical time scale associated with a Langevin process, $\tau_\text{L}=\gamma_\text{L}^{-1} n_\text{a}^{-1}$, where $n_a$ is the atom density. In particular, we use experimentally realistic densities for Rb and Li, given by $n_{\text{Rb}} = 10^{20}\,$m$^{-3}$~\cite{Mohammadi2021} and  $n_{\text{Li}}=10^{18}\,$m$^{-3}$~\cite{Weckesser:2021}, respectively, and the results are depicted as the dashed lines in the lower panel of Fig.~\ref{fig:masses}. For the lowest density considered, HT and PT cases present a complex lifetime much shorter than the Langevin time. On the contrary, for $n_{\text{Rb}}$ in the HT scenario, the complex lifetime is longer than the Langevin time. In that case, there is a high probability that a third body collides with the complex before decaying, leading to the formation of a stable molecule via three-body recombination. Besides, when the ion is held in a PT, the complex lifetime is shorter than the Langevin time, thus, suppressing three-body recombination reactions, as we discuss below. 

\begin{figure}
	\centering
	\includegraphics[width=\linewidth]{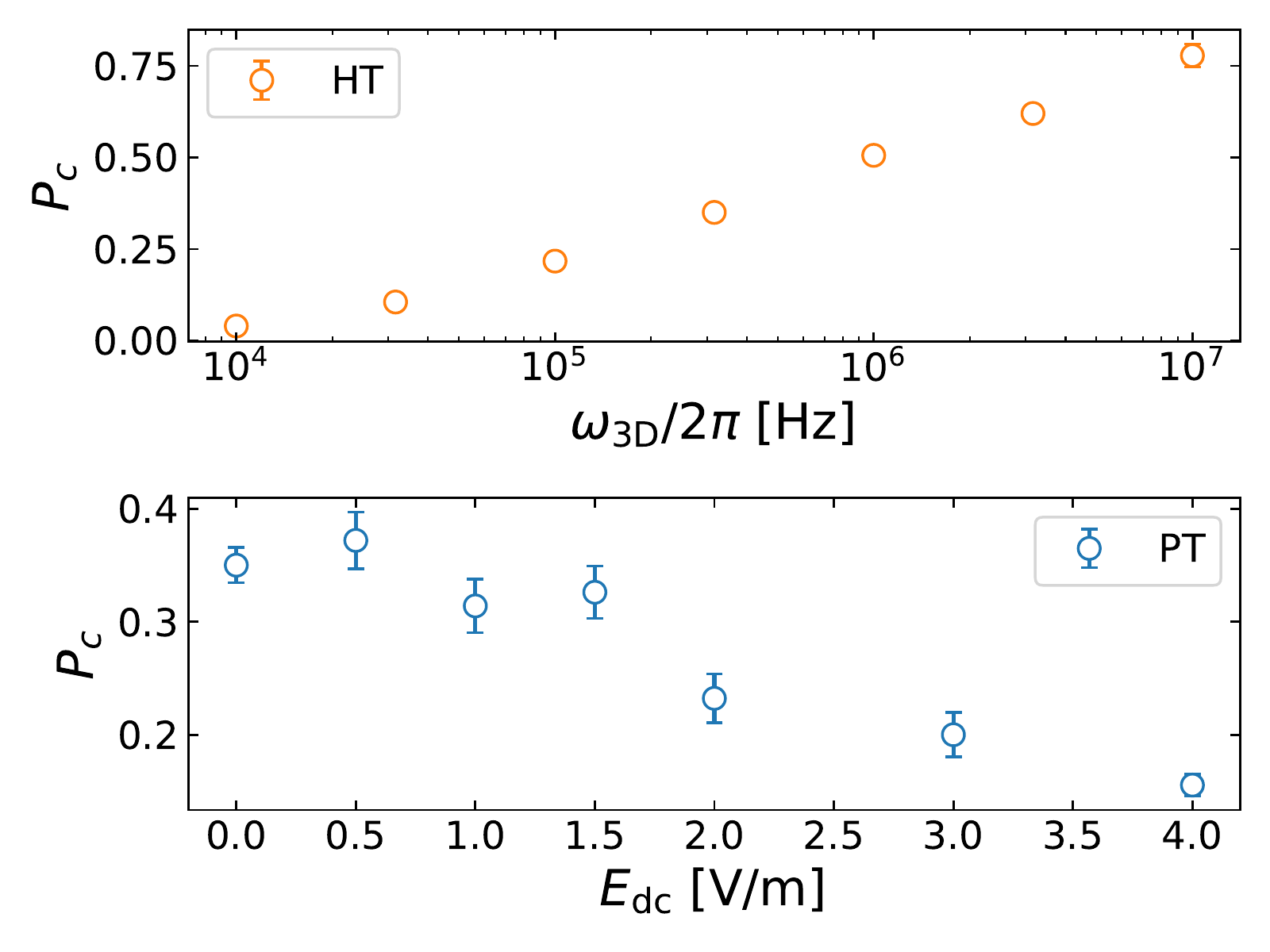}
 	\caption{Upper panel: Complex formation probability in a harmonic trap (HT) as a function of the trap frequency. Lower panel: Complex formation in a Paul trap (PT) and influence of the ion micromotion. Micromotion energy is increased by applying an electric field $E_\mathrm{dc}$ which shifts the ion from the rf-zero node.}
 	\label{fig:HT}
\end{figure}

Fig~\ref{fig:HT} displays our results for the complex formation probability for Yb$^+$ ($T_\mathrm{ion}=100\,\mu$K) - $^6$Li ($T_\text{a}=2\,\mu$K) collisions. The upper panel refers to HT case for the trap frequencies $\omega_x\approx\omega_y\approx\omega_z\approx\omega_{3\mathrm{D}}$ in a range of $10\,$kHz to $10\,$MHz. With the used parameter set, we find a significant effect of the trap frequency leading to a variation of the complex formation probability between 10\,\% and 77\,\%. On the other hand, the lower panel, referring to the PT case, shows a significant influence of the micromotion on complex formation. In particular, we assume an ideal PT but adding an additional electric dc field $E_\mathrm{dc}$ to push the ion from the center of the rf-field to increase the micromotion amplitude. As a result, a general trend is noticeable: larger $E_\mathrm{dc}$ fields lead to a lower complex formation probability. In particular, we observe that the probability of complex formation remains mainly the same for $E_\mathrm{dc}\lesssim 1.5\,$V/m. However, adding a $E_\mathrm{dc}\approx 2\,$V/m the $P_\text{c}$ is reduced by $50\,$\% compared to the ideal case, suggesting the the existence of a threshold for $E_\mathrm{dc}\approx 2\,$V/m.

\begin{figure}
	\centering
	\includegraphics[width=\linewidth]{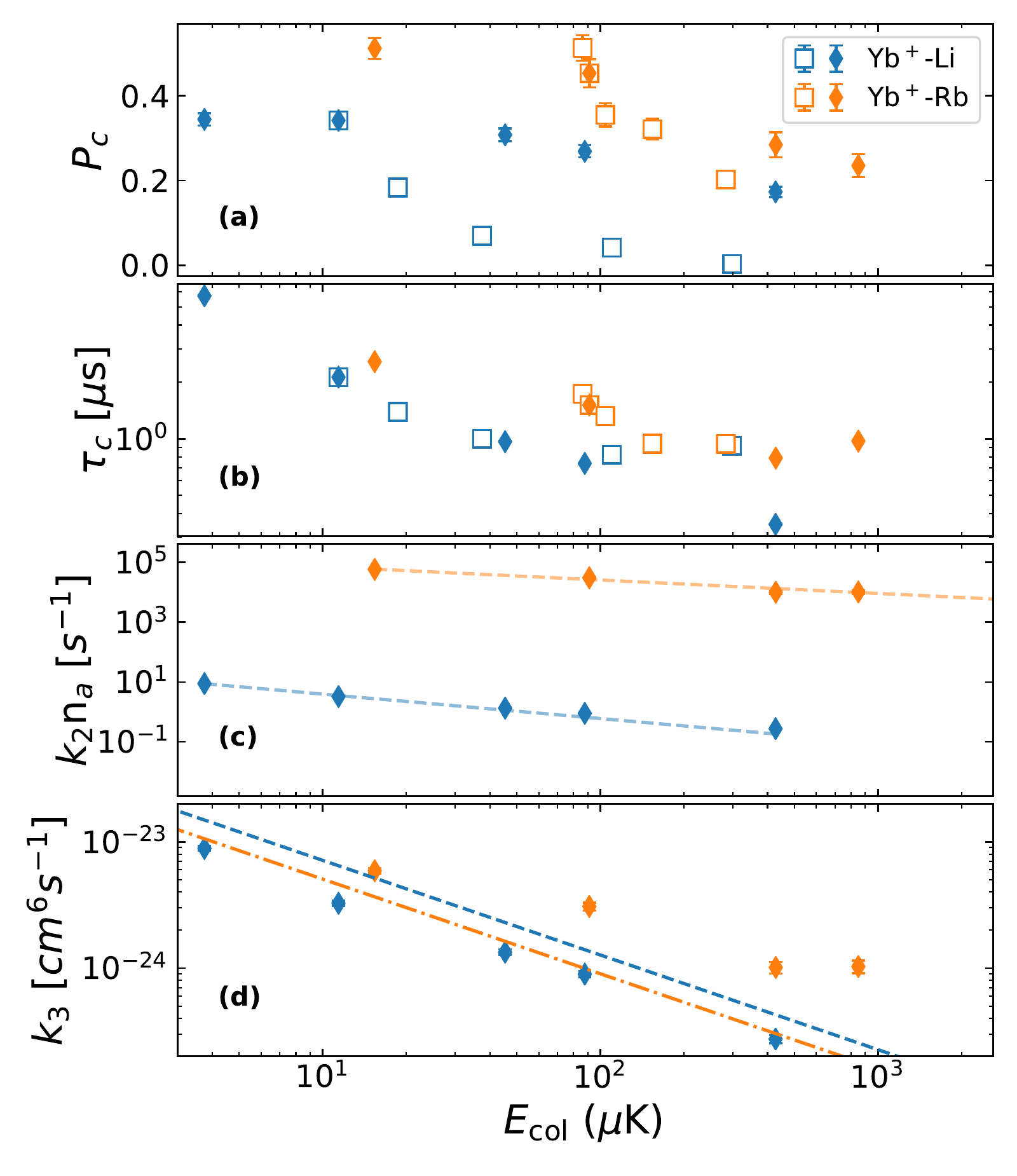}
 	\caption{Atom-ion complexes in the Paul trap versus collision energy for the combinations Yb$^+$-Li (blue) and Yb$^+$-Rb (orange). The collision energy is varied by changing the atom energy ($\square$) or the ion energy ($\blacklozenge$). In a) the complex formation probability and in b) the complex lifetime is shown. Panel c) shows $k_2$, assuming $n_\mathrm{Li}=10^{18}$m$^{-3}$ and $n_\mathrm{Rb}=10^{20}$m$^{-3}$ with power-law fits (dashed lines). In Panel d) the numerical results of $k_3$ are shown together with the solution of the analytic expression for $k_3$ in absence of a trap (dashed lines). }
 	\label{fig4}
\end{figure}

From now on, we will focus on the PT scenario. First, by looking into the role of collision energy on the probability of complex formation and its lifetime. The results are shown in Fig.~\ref{fig4} for Yb$^+$-Li (blue) and Yb$^+$-Rb (orange). We notice, as expected, that lower collision energies lead to more complexes compared to the case of higher collision energies. Similarly, the same observation holds for $\tau_{\text{c}}$. In addition, we notice that for similar collision energies, the impact of modifying the ion energy ($\blacklozenge$) is different from the atom one ($\square$) on $P_{\text{c}}$ and $\tau_{\text{c}}$, which is due to the presence of the trap.

\paragraph{Three-body recombination}

Ion-atom-atom three-body recombination is a termolecular reaction process in which three free particles react into a molecule plus a free atom,
\ce{ A+ + B  + B ->[k_3] AB+ + B }, where $k_3$ stands for the three-body recombination rate. Three-body processes can be viewed as the result of two bi-molecular processes. A prime example of this approach is the well-known stabilization and Chaperon mechanism relevant for ozone formation~\cite{ozone1,ozone2}, or the Lindemann-Hinshelwood mechanism, known as the Roberts- Bernstein-Curtiss mechanism in the three neutral atom case~\cite{RBCmechanism,RBCmechanism2}. In our case, the bi-molecular processes are 

\begin{align}
\ce{ A+ + B <-->[k_2][k_{diss}] (AB+)^* }\\
\ce{ (AB+)^* + B ->[k_{est}] AB+ + B },
\end{align}
where $k_2$ denotes the rate of formation of (AB$^+$)$^*$ complexes, $k_{\text{diss}}$ stands for its dissociation rate and $k_{\text{est}}$ refers to the stabilization rate due to a collision with a third body. Indeed, assuming that the production of complexes reaches a steady state, we find that the three-body recombination rate is given by

\begin{equation}\label{eq7}
    k_3(E_{\text{col}})=\frac{k_2(E_{\text{col}})k_{\text{est}}(E_{\text{col}})}{k_\text{diss}(E_{\text{col}})+k_{\text{est}}(E_{\text{col}})[\text{B}]},
\end{equation}
where [B] is the number density of particle B.

In the limit of high atomic density, the three-body recombination rate reads as

\begin{equation}\label{eq8}
    k_3(E_{\text{col}})=\frac{k_2(E_{\text{col}})}{[\text{B}]}.
\end{equation}
In other words, every atom-ion complex will lead to the formation of a stable molecular ion. Then, the three-body recombination rate is directly proportional to $k_2$ and Eq.~(\ref{eq8}) describes an upper bound for the ion-atom-atom three-body recombination rate in the presence of a trap. 

We calculate the formation rate of atom-ion complexes, $k_2$ as
\begin{equation}
\label{eq9}
	k_{2} = k_\text{L} \frac{P_\text{c}\tau_\text{c}}{\tau_\text{L}} =  (2\pi)^2 n_\text{a} \frac{\alpha}{\mu} P_\text{c}\tau_\text{c}, 
\end{equation}
where we take $P_\text{c}\tau_\text{c}/\tau_\text{L}$ as the probability of complex formation during a Langevin collision. Thereby, $P_c$ takes into account, that not every Langevin collision leads to a complex. The results for $k_2$ as a function of the collision energy for a single Yb$^+$ ion colliding with Rb and Li atoms is shown in panel (c) of Fig.\ref{fig4}. We notice that the complex rate formation depends drastically on the mass of the atom, in agreement with Fig.~\ref{fig:masses}. Indeed, surprisingly enough, we identify that $k_2$ shows a different energy-dependent behavior based on the mass of the colliding atom. In particular, after fitting the formation rate to a function $E_{\text{col}}^{\beta}$, we find $\beta=$ -0.82 (-0.45) for Li (Rb). This behaviour can only be explained via the effect of the trap on the ion since the energy dependence should be dominated by the long-range tail of the atom-ion potential, which has the same dependency for the two cases under consideration.  

Once $k_2$ is computed, we can calculate the ion-atom-atom three-body recombination rate in the presence of a trap. The results are shown in panel (d) of Fig.\ref{fig4}, where it is noticed that Yb$^+$-Li shows a slower rate than Yb$^+$-Rb, as expected based on the complex formation rate (see panel (c) Fig.\ref{fig4}) and the mass of the atom. Meanwhile, the energy dependency of the rate is different for different atoms in stark contrast with free-trap collisions, in which $k_3\propto E_{\text{col}}^{-3/4}$~\cite{perezrios2015,kruekow2016,perezrios2018}, depicted as the dashed lines in panel (d) of Fig.\ref{fig4}. In particular, our results for Yb$^+$-Li agree fairly well with the free-field prediction (dashed-blue line). However, Yb$^+$-Rb shows a much larger rate and a less steep power-law than in the free-field case (dashed-orange line).

%%%%%%%%%%%%%%%%%%%%%%%%%%%%%%%%%%%%%%%%%%%%%%%%%
\paragraph{Conclusion}
This study predicts the existence of atom-ion complex formation in traps regardless of the nature of the trap. These results may look surprising under previous studies, in which the formation of complexes was mainly attributed to the time-dependent nature of the trapping potential~\cite{Cetina:2012, meir2017}. Indeed, our findings show that it is possible to control three-body processes via trap parameters, collision energy, and atomic species.

In the case of an ion held in a Paul trap, we have shown that the probability of the atom-ion complex formation can be readily controlled by increasing the micromotion amplitude via an additional electric field. Additionally, we identify that heavier atoms lead to a more significant probability of complex formation, although they have a similar complex lifetime to light atoms. Furthermore, we predict the ion-atom-atom three-body recombination rate in the presence of the trap assuming a large atomic density, where we notice a significant effect of the atomic mass on the energy-dependent three-body recombination rate. This behavior is due to the presence of the trap and can not be rationalized in light of direct three-body recombination reactions in free space. On the other hand, we have shown that the static confining fields of ion optical traps may affect the stability of the ion when brought in contact with a given atomic species~\cite{schneider:2010} due to a probable enhancement of three-body losses.

While working on this manuscript we became aware of a recent experiment reporting experimental evidence of trap assisted complexes~\cite{Katz2022,Pinkas2022}.

\section*{Acknowledgements}
We thank Marko Cetina, Or Katz, and Seth Rittenhouse for fruitful discussions. J.P.-R. thanks the support of the Simons Foundation. This work was supported by the Dutch Research Council Start-up grant 740.018.008 (H.H. and R.G.), Vrije Programma 680.92.18.05 (E.T., R.G. and J.P.-R.) and Quantum Software Consortium programme 024.003.037 (A.S.N.). 
%\bibliography{hh_complex}%, mainNotes}

\begin{thebibliography}{40}%
\makeatletter
\providecommand \@ifxundefined [1]{%
 \@ifx{#1\undefined}
}%
\providecommand \@ifnum [1]{%
 \ifnum #1\expandafter \@firstoftwo
 \else \expandafter \@secondoftwo
 \fi
}%
\providecommand \@ifx [1]{%
 \ifx #1\expandafter \@firstoftwo
 \else \expandafter \@secondoftwo
 \fi
}%
\providecommand \natexlab [1]{#1}%
\providecommand \enquote  [1]{``#1''}%
\providecommand \bibnamefont  [1]{#1}%
\providecommand \bibfnamefont [1]{#1}%
\providecommand \citenamefont [1]{#1}%
\providecommand \href@noop [0]{\@secondoftwo}%
\providecommand \href [0]{\begingroup \@sanitize@url \@href}%
\providecommand \@href[1]{\@@startlink{#1}\@@href}%
\providecommand \@@href[1]{\endgroup#1\@@endlink}%
\providecommand \@sanitize@url [0]{\catcode `\\12\catcode `\$12\catcode
  `\&12\catcode `\#12\catcode `\^12\catcode `\_12\catcode `\%12\relax}%
\providecommand \@@startlink[1]{}%
\providecommand \@@endlink[0]{}%
\providecommand \url  [0]{\begingroup\@sanitize@url \@url }%
\providecommand \@url [1]{\endgroup\@href {#1}{\urlprefix }}%
\providecommand \urlprefix  [0]{URL }%
\providecommand \Eprint [0]{\href }%
\providecommand \doibase [0]{http://dx.doi.org/}%
\providecommand \selectlanguage [0]{\@gobble}%
\providecommand \bibinfo  [0]{\@secondoftwo}%
\providecommand \bibfield  [0]{\@secondoftwo}%
\providecommand \translation [1]{[#1]}%
\providecommand \BibitemOpen [0]{}%
\providecommand \bibitemStop [0]{}%
\providecommand \bibitemNoStop [0]{.\EOS\space}%
\providecommand \EOS [0]{\spacefactor3000\relax}%
\providecommand \BibitemShut  [1]{\csname bibitem#1\endcsname}%
\let\auto@bib@innerbib\@empty
%</preamble>
\bibitem [{\citenamefont {Troe}(1994)}]{Troe1994}%
  \BibitemOpen
  \bibfield  {author} {\bibinfo {author} {\bibfnamefont {J.}~\bibnamefont
  {Troe}},\ }\href {\doibase 10.1039/ft9949002303} {\bibfield  {journal}
  {\bibinfo  {journal} {Journal of the Chemical Society, Faraday Transactions}\
  }\textbf {\bibinfo {volume} {90}},\ \bibinfo {pages} {2303} (\bibinfo {year}
  {1994})}\BibitemShut {NoStop}%
\bibitem [{\citenamefont {Su}\ \emph {et~al.}(2013)\citenamefont {Su},
  \citenamefont {Huang}, \citenamefont {Witek},\ and\ \citenamefont
  {Lee}}]{Su2013}%
  \BibitemOpen
  \bibfield  {author} {\bibinfo {author} {\bibfnamefont {Y.-T.}\ \bibnamefont
  {Su}}, \bibinfo {author} {\bibfnamefont {Y.-H.}\ \bibnamefont {Huang}},
  \bibinfo {author} {\bibfnamefont {H.~A.}\ \bibnamefont {Witek}}, \ and\
  \bibinfo {author} {\bibfnamefont {Y.-P.}\ \bibnamefont {Lee}},\ }\href
  {\doibase 10.1126/science.1234369} {\bibfield  {journal} {\bibinfo  {journal}
  {Science}\ }\textbf {\bibinfo {volume} {340}},\ \bibinfo {pages} {174}
  (\bibinfo {year} {2013})}\BibitemShut {NoStop}%
\bibitem [{\citenamefont {Bjork}\ \emph {et~al.}(2016)\citenamefont {Bjork},
  \citenamefont {Bui}, \citenamefont {Heckl}, \citenamefont {Changala},
  \citenamefont {Spaun}, \citenamefont {Heu}, \citenamefont {Follman},
  \citenamefont {Deutsch}, \citenamefont {Cole}, \citenamefont {Aspelmeyer},
  \citenamefont {Okumura},\ and\ \citenamefont {Ye}}]{Bjork2016}%
  \BibitemOpen
  \bibfield  {author} {\bibinfo {author} {\bibfnamefont {B.~J.}\ \bibnamefont
  {Bjork}}, \bibinfo {author} {\bibfnamefont {T.~Q.}\ \bibnamefont {Bui}},
  \bibinfo {author} {\bibfnamefont {O.~H.}\ \bibnamefont {Heckl}}, \bibinfo
  {author} {\bibfnamefont {P.~B.}\ \bibnamefont {Changala}}, \bibinfo {author}
  {\bibfnamefont {B.}~\bibnamefont {Spaun}}, \bibinfo {author} {\bibfnamefont
  {P.}~\bibnamefont {Heu}}, \bibinfo {author} {\bibfnamefont {D.}~\bibnamefont
  {Follman}}, \bibinfo {author} {\bibfnamefont {C.}~\bibnamefont {Deutsch}},
  \bibinfo {author} {\bibfnamefont {G.~D.}\ \bibnamefont {Cole}}, \bibinfo
  {author} {\bibfnamefont {M.}~\bibnamefont {Aspelmeyer}}, \bibinfo {author}
  {\bibfnamefont {M.}~\bibnamefont {Okumura}}, \ and\ \bibinfo {author}
  {\bibfnamefont {J.}~\bibnamefont {Ye}},\ }\href {\doibase
  10.1126/science.aag1862} {\bibfield  {journal} {\bibinfo  {journal}
  {Science}\ }\textbf {\bibinfo {volume} {354}},\ \bibinfo {pages} {444}
  (\bibinfo {year} {2016})}\BibitemShut {NoStop}%
\bibitem [{\citenamefont {Hu}\ \emph {et~al.}(2019)\citenamefont {Hu},
  \citenamefont {Liu}, \citenamefont {Grimes}, \citenamefont {Lin},
  \citenamefont {Gheorghe}, \citenamefont {Vexiau}, \citenamefont
  {Bouloufa-Maafa}, \citenamefont {Dulieu}, \citenamefont {Rosenband},\ and\
  \citenamefont {Ni}}]{Hu2019}%
  \BibitemOpen
  \bibfield  {author} {\bibinfo {author} {\bibfnamefont {M.-G.}\ \bibnamefont
  {Hu}}, \bibinfo {author} {\bibfnamefont {Y.}~\bibnamefont {Liu}}, \bibinfo
  {author} {\bibfnamefont {D.~D.}\ \bibnamefont {Grimes}}, \bibinfo {author}
  {\bibfnamefont {Y.-W.}\ \bibnamefont {Lin}}, \bibinfo {author} {\bibfnamefont
  {A.~H.}\ \bibnamefont {Gheorghe}}, \bibinfo {author} {\bibfnamefont
  {R.}~\bibnamefont {Vexiau}}, \bibinfo {author} {\bibfnamefont
  {N.}~\bibnamefont {Bouloufa-Maafa}}, \bibinfo {author} {\bibfnamefont
  {O.}~\bibnamefont {Dulieu}}, \bibinfo {author} {\bibfnamefont
  {T.}~\bibnamefont {Rosenband}}, \ and\ \bibinfo {author} {\bibfnamefont
  {K.-K.}\ \bibnamefont {Ni}},\ }\href {\doibase 10.1126/science.aay9531}
  {\bibfield  {journal} {\bibinfo  {journal} {Science}\ }\textbf {\bibinfo
  {volume} {366}},\ \bibinfo {pages} {1111} (\bibinfo {year}
  {2019})}\BibitemShut {NoStop}%
\bibitem [{\citenamefont {Nichols}\ \emph {et~al.}(2022)\citenamefont
  {Nichols}, \citenamefont {Liu}, \citenamefont {Zhu}, \citenamefont {Hu},
  \citenamefont {Liu},\ and\ \citenamefont {Ni}}]{observation}%
  \BibitemOpen
  \bibfield  {author} {\bibinfo {author} {\bibfnamefont {M.~A.}\ \bibnamefont
  {Nichols}}, \bibinfo {author} {\bibfnamefont {Y.-X.}\ \bibnamefont {Liu}},
  \bibinfo {author} {\bibfnamefont {L.}~\bibnamefont {Zhu}}, \bibinfo {author}
  {\bibfnamefont {M.-G.}\ \bibnamefont {Hu}}, \bibinfo {author} {\bibfnamefont
  {Y.}~\bibnamefont {Liu}}, \ and\ \bibinfo {author} {\bibfnamefont {K.-K.}\
  \bibnamefont {Ni}},\ }\href {\doibase 10.1103/PhysRevX.12.011049} {\bibfield
  {journal} {\bibinfo  {journal} {Phys. Rev. X}\ }\textbf {\bibinfo {volume}
  {12}},\ \bibinfo {pages} {011049} (\bibinfo {year} {2022})}\BibitemShut
  {NoStop}%
\bibitem [{\citenamefont {Liu}\ \emph {et~al.}(2020)\citenamefont {Liu},
  \citenamefont {Hu}, \citenamefont {Nichols}, \citenamefont {Grimes},
  \citenamefont {Karman}, \citenamefont {Guo},\ and\ \citenamefont
  {Ni}}]{observation_bis}%
  \BibitemOpen
  \bibfield  {author} {\bibinfo {author} {\bibfnamefont {Y.}~\bibnamefont
  {Liu}}, \bibinfo {author} {\bibfnamefont {M.-G.}\ \bibnamefont {Hu}},
  \bibinfo {author} {\bibfnamefont {M.~A.}\ \bibnamefont {Nichols}}, \bibinfo
  {author} {\bibfnamefont {D.~D.}\ \bibnamefont {Grimes}}, \bibinfo {author}
  {\bibfnamefont {T.}~\bibnamefont {Karman}}, \bibinfo {author} {\bibfnamefont
  {H.}~\bibnamefont {Guo}}, \ and\ \bibinfo {author} {\bibfnamefont {K.-K.}\
  \bibnamefont {Ni}},\ }\href {\doibase 10.1038/s41567-020-0968-8} {\bibfield
  {journal} {\bibinfo  {journal} {Nature Physics}\ }\textbf {\bibinfo {volume}
  {16}},\ \bibinfo {pages} {1132} (\bibinfo {year} {2020})}\BibitemShut
  {NoStop}%
\bibitem [{\citenamefont {Gersema}\ \emph {et~al.}(2021)\citenamefont
  {Gersema}, \citenamefont {Voges}, \citenamefont {Meyer~zum Alten~Borgloh},
  \citenamefont {Koch}, \citenamefont {Hartmann}, \citenamefont {Zenesini},
  \citenamefont {Ospelkaus}, \citenamefont {Lin}, \citenamefont {He},\ and\
  \citenamefont {Wang}}]{observationHamburg}%
  \BibitemOpen
  \bibfield  {author} {\bibinfo {author} {\bibfnamefont {P.}~\bibnamefont
  {Gersema}}, \bibinfo {author} {\bibfnamefont {K.~K.}\ \bibnamefont {Voges}},
  \bibinfo {author} {\bibfnamefont {M.}~\bibnamefont {Meyer~zum
  Alten~Borgloh}}, \bibinfo {author} {\bibfnamefont {L.}~\bibnamefont {Koch}},
  \bibinfo {author} {\bibfnamefont {T.}~\bibnamefont {Hartmann}}, \bibinfo
  {author} {\bibfnamefont {A.}~\bibnamefont {Zenesini}}, \bibinfo {author}
  {\bibfnamefont {S.}~\bibnamefont {Ospelkaus}}, \bibinfo {author}
  {\bibfnamefont {J.}~\bibnamefont {Lin}}, \bibinfo {author} {\bibfnamefont
  {J.}~\bibnamefont {He}}, \ and\ \bibinfo {author} {\bibfnamefont
  {D.}~\bibnamefont {Wang}},\ }\href {\doibase 10.1103/PhysRevLett.127.163401}
  {\bibfield  {journal} {\bibinfo  {journal} {Phys. Rev. Lett.}\ }\textbf
  {\bibinfo {volume} {127}},\ \bibinfo {pages} {163401} (\bibinfo {year}
  {2021})}\BibitemShut {NoStop}%
\bibitem [{\citenamefont {Gregory}\ \emph {et~al.}(2019)\citenamefont
  {Gregory}, \citenamefont {Frye}, \citenamefont {Blackmore}, \citenamefont
  {Bridge}, \citenamefont {Sawant}, \citenamefont {Hutson},\ and\ \citenamefont
  {Cornish}}]{observationRbCs}%
  \BibitemOpen
  \bibfield  {author} {\bibinfo {author} {\bibfnamefont {P.~D.}\ \bibnamefont
  {Gregory}}, \bibinfo {author} {\bibfnamefont {M.~D.}\ \bibnamefont {Frye}},
  \bibinfo {author} {\bibfnamefont {J.~A.}\ \bibnamefont {Blackmore}}, \bibinfo
  {author} {\bibfnamefont {E.~M.}\ \bibnamefont {Bridge}}, \bibinfo {author}
  {\bibfnamefont {R.}~\bibnamefont {Sawant}}, \bibinfo {author} {\bibfnamefont
  {J.~M.}\ \bibnamefont {Hutson}}, \ and\ \bibinfo {author} {\bibfnamefont
  {S.~L.}\ \bibnamefont {Cornish}},\ }\href {\doibase
  10.1038/s41467-019-11033-y} {\bibfield  {journal} {\bibinfo  {journal}
  {Nature Communications}\ }\textbf {\bibinfo {volume} {10}},\ \bibinfo {pages}
  {3104} (\bibinfo {year} {2019})}\BibitemShut {NoStop}%
\bibitem [{\citenamefont {Zipkes}\ \emph {et~al.}(2010)\citenamefont {Zipkes},
  \citenamefont {Palzer}, \citenamefont {Ratschbacher}, \citenamefont {Sias},\
  and\ \citenamefont {K\"ohl}}]{zipkes:2010b}%
  \BibitemOpen
  \bibfield  {author} {\bibinfo {author} {\bibfnamefont {C.}~\bibnamefont
  {Zipkes}}, \bibinfo {author} {\bibfnamefont {S.}~\bibnamefont {Palzer}},
  \bibinfo {author} {\bibfnamefont {L.}~\bibnamefont {Ratschbacher}}, \bibinfo
  {author} {\bibfnamefont {C.}~\bibnamefont {Sias}}, \ and\ \bibinfo {author}
  {\bibfnamefont {M.}~\bibnamefont {K\"ohl}},\ }\href {\doibase
  10.1103/PhysRevLett.105.133201} {\bibfield  {journal} {\bibinfo  {journal}
  {Phys. Rev. Lett.}\ }\textbf {\bibinfo {volume} {105}},\ \bibinfo {pages}
  {133201} (\bibinfo {year} {2010})}\BibitemShut {NoStop}%
\bibitem [{\citenamefont {Schmid}\ \emph {et~al.}(2010)\citenamefont {Schmid},
  \citenamefont {H\"arter},\ and\ \citenamefont {Denschlag}}]{Schmid:2010}%
  \BibitemOpen
  \bibfield  {author} {\bibinfo {author} {\bibfnamefont {S.}~\bibnamefont
  {Schmid}}, \bibinfo {author} {\bibfnamefont {A.}~\bibnamefont {H\"arter}}, \
  and\ \bibinfo {author} {\bibfnamefont {J.~H.}\ \bibnamefont {Denschlag}},\
  }\href {\doibase 10.1103/PhysRevLett.105.133202} {\bibfield  {journal}
  {\bibinfo  {journal} {Phys. Rev. Lett.}\ }\textbf {\bibinfo {volume} {105}},\
  \bibinfo {pages} {133202} (\bibinfo {year} {2010})}\BibitemShut {NoStop}%
\bibitem [{\citenamefont {Ravi}\ \emph {et~al.}(2012)\citenamefont {Ravi},
  \citenamefont {Lee}, \citenamefont {Sharma}, \citenamefont {Werth},\ and\
  \citenamefont {Rangwala}}]{Ravi:2012}%
  \BibitemOpen
  \bibfield  {author} {\bibinfo {author} {\bibfnamefont {K.}~\bibnamefont
  {Ravi}}, \bibinfo {author} {\bibfnamefont {S.}~\bibnamefont {Lee}}, \bibinfo
  {author} {\bibfnamefont {A.}~\bibnamefont {Sharma}}, \bibinfo {author}
  {\bibfnamefont {G.}~\bibnamefont {Werth}}, \ and\ \bibinfo {author}
  {\bibfnamefont {S.~A.}\ \bibnamefont {Rangwala}},\ }\href {\doibase
  10.1038/ncomms2131} {\bibfield  {journal} {\bibinfo  {journal} {Nat.
  Commun.}\ }\textbf {\bibinfo {volume} {3}},\ \bibinfo {pages} {1126}
  (\bibinfo {year} {2012})}\BibitemShut {NoStop}%
\bibitem [{\citenamefont {Haze}\ \emph {et~al.}(2015)\citenamefont {Haze},
  \citenamefont {Saito}, \citenamefont {Fujinaga},\ and\ \citenamefont
  {Mukaiyama}}]{haze:2015}%
  \BibitemOpen
  \bibfield  {author} {\bibinfo {author} {\bibfnamefont {S.}~\bibnamefont
  {Haze}}, \bibinfo {author} {\bibfnamefont {R.}~\bibnamefont {Saito}},
  \bibinfo {author} {\bibfnamefont {M.}~\bibnamefont {Fujinaga}}, \ and\
  \bibinfo {author} {\bibfnamefont {T.}~\bibnamefont {Mukaiyama}},\ }\href
  {\doibase 10.1103/PhysRevA.91.032709} {\bibfield  {journal} {\bibinfo
  {journal} {Phys. Rev. A}\ }\textbf {\bibinfo {volume} {91}},\ \bibinfo
  {pages} {032709} (\bibinfo {year} {2015})}\BibitemShut {NoStop}%
\bibitem [{\citenamefont {Saito}\ \emph {et~al.}(2017)\citenamefont {Saito},
  \citenamefont {Haze}, \citenamefont {Sasakawa}, \citenamefont {Nakai},
  \citenamefont {Raoult}, \citenamefont {Silva}, \citenamefont {Dulieu},\ and\
  \citenamefont {Mukaiyama}}]{Saito:2017}%
  \BibitemOpen
  \bibfield  {author} {\bibinfo {author} {\bibfnamefont {R.}~\bibnamefont
  {Saito}}, \bibinfo {author} {\bibfnamefont {S.}~\bibnamefont {Haze}},
  \bibinfo {author} {\bibfnamefont {M.}~\bibnamefont {Sasakawa}}, \bibinfo
  {author} {\bibfnamefont {R.}~\bibnamefont {Nakai}}, \bibinfo {author}
  {\bibfnamefont {M.}~\bibnamefont {Raoult}}, \bibinfo {author} {\bibfnamefont
  {H.~D.}\ \bibnamefont {Silva}}, \bibinfo {author} {\bibfnamefont
  {O.}~\bibnamefont {Dulieu}}, \ and\ \bibinfo {author} {\bibfnamefont
  {T.}~\bibnamefont {Mukaiyama}},\ }\href {\doibase 10.1103/PhysRevA.95.032709}
  {\bibfield  {journal} {\bibinfo  {journal} {Phys.~Rev.~A}\ }\textbf {\bibinfo
  {volume} {95}},\ \bibinfo {pages} {032709} (\bibinfo {year}
  {2017})}\BibitemShut {NoStop}%
\bibitem [{\citenamefont {Joger}\ \emph {et~al.}(2017)\citenamefont {Joger},
  \citenamefont {F\"urst}, \citenamefont {Ewald}, \citenamefont {Feldker},
  \citenamefont {Tomza},\ and\ \citenamefont {Gerritsma}}]{Joger:2017}%
  \BibitemOpen
  \bibfield  {author} {\bibinfo {author} {\bibfnamefont {J.}~\bibnamefont
  {Joger}}, \bibinfo {author} {\bibfnamefont {H.}~\bibnamefont {F\"urst}},
  \bibinfo {author} {\bibfnamefont {N.}~\bibnamefont {Ewald}}, \bibinfo
  {author} {\bibfnamefont {T.}~\bibnamefont {Feldker}}, \bibinfo {author}
  {\bibfnamefont {M.}~\bibnamefont {Tomza}}, \ and\ \bibinfo {author}
  {\bibfnamefont {R.}~\bibnamefont {Gerritsma}},\ }\href {\doibase
  10.1103/PhysRevA.96.030703} {\bibfield  {journal} {\bibinfo  {journal} {Phys.
  Rev. A}\ }\textbf {\bibinfo {volume} {96}},\ \bibinfo {pages} {030703(R)}
  (\bibinfo {year} {2017})}\BibitemShut {NoStop}%
\bibitem [{\citenamefont {Sikorsky}\ \emph {et~al.}(2018)\citenamefont
  {Sikorsky}, \citenamefont {Meir}, \citenamefont {Ben-shlomi}, \citenamefont
  {Akerman},\ and\ \citenamefont {Ozeri}}]{Sikorsky:2018}%
  \BibitemOpen
  \bibfield  {author} {\bibinfo {author} {\bibfnamefont {T.}~\bibnamefont
  {Sikorsky}}, \bibinfo {author} {\bibfnamefont {Z.}~\bibnamefont {Meir}},
  \bibinfo {author} {\bibfnamefont {R.}~\bibnamefont {Ben-shlomi}}, \bibinfo
  {author} {\bibfnamefont {N.}~\bibnamefont {Akerman}}, \ and\ \bibinfo
  {author} {\bibfnamefont {R.}~\bibnamefont {Ozeri}},\ }\href {\doibase
  10.1038/s41467-018-03373-y} {\bibfield  {journal} {\bibinfo  {journal} {Nat.
  Commun.}\ }\textbf {\bibinfo {volume} {9}},\ \bibinfo {pages} {920} (\bibinfo
  {year} {2018})}\BibitemShut {NoStop}%
\bibitem [{\citenamefont {Li}\ \emph {et~al.}(2020)\citenamefont {Li},
  \citenamefont {Jyothi}, \citenamefont {Li}, \citenamefont {K{\l}os},
  \citenamefont {Petrov}, \citenamefont {Brown},\ and\ \citenamefont
  {Kotochigova}}]{Li2020}%
  \BibitemOpen
  \bibfield  {author} {\bibinfo {author} {\bibfnamefont {H.}~\bibnamefont
  {Li}}, \bibinfo {author} {\bibfnamefont {S.}~\bibnamefont {Jyothi}}, \bibinfo
  {author} {\bibfnamefont {M.}~\bibnamefont {Li}}, \bibinfo {author}
  {\bibfnamefont {J.}~\bibnamefont {K{\l}os}}, \bibinfo {author} {\bibfnamefont
  {A.}~\bibnamefont {Petrov}}, \bibinfo {author} {\bibfnamefont {K.~R.}\
  \bibnamefont {Brown}}, \ and\ \bibinfo {author} {\bibfnamefont
  {S.}~\bibnamefont {Kotochigova}},\ }\href {\doibase 10.1039/d0cp01131b}
  {\bibfield  {journal} {\bibinfo  {journal} {Phys. Chem. Chem. Phys.}\
  }\textbf {\bibinfo {volume} {22}},\ \bibinfo {pages} {10870} (\bibinfo {year}
  {2020})}\BibitemShut {NoStop}%
\bibitem [{\citenamefont {Mahdian}\ \emph {et~al.}(2021)\citenamefont
  {Mahdian}, \citenamefont {Krükow},\ and\ \citenamefont
  {Denschlag}}]{mahdian2021}%
  \BibitemOpen
  \bibfield  {author} {\bibinfo {author} {\bibfnamefont {A.}~\bibnamefont
  {Mahdian}}, \bibinfo {author} {\bibfnamefont {A.}~\bibnamefont {Krükow}}, \
  and\ \bibinfo {author} {\bibfnamefont {J.~H.}\ \bibnamefont {Denschlag}},\
  }\href {\doibase 10.1088/1367-2630/ac0575} {\bibfield  {journal} {\bibinfo
  {journal} {New J.~Phys.}\ }\textbf {\bibinfo {volume} {23}},\ \bibinfo
  {pages} {065008} (\bibinfo {year} {2021})}\BibitemShut {NoStop}%
\bibitem [{\citenamefont {Mohammadi}\ \emph {et~al.}(2021)\citenamefont
  {Mohammadi}, \citenamefont {Krükow}, \citenamefont {Mahdian}, \citenamefont
  {Dei{\ss}}, \citenamefont {P{\'{e}}rez-R{\'{\i}}os}, \citenamefont
  {da~Silva}, \citenamefont {Raoult}, \citenamefont {Dulieu},\ and\
  \citenamefont {Denschlag}}]{Mohammadi2021}%
  \BibitemOpen
  \bibfield  {author} {\bibinfo {author} {\bibfnamefont {A.}~\bibnamefont
  {Mohammadi}}, \bibinfo {author} {\bibfnamefont {A.}~\bibnamefont {Krükow}},
  \bibinfo {author} {\bibfnamefont {A.}~\bibnamefont {Mahdian}}, \bibinfo
  {author} {\bibfnamefont {M.}~\bibnamefont {Dei{\ss}}}, \bibinfo {author}
  {\bibfnamefont {J.}~\bibnamefont {P{\'{e}}rez-R{\'{\i}}os}}, \bibinfo
  {author} {\bibfnamefont {H.}~\bibnamefont {da~Silva}}, \bibinfo {author}
  {\bibfnamefont {M.}~\bibnamefont {Raoult}}, \bibinfo {author} {\bibfnamefont
  {O.}~\bibnamefont {Dulieu}}, \ and\ \bibinfo {author} {\bibfnamefont {J.~H.}\
  \bibnamefont {Denschlag}},\ }\href
  {https://doi.org/10.1103/PhysRevResearch.3.013196} {\bibfield  {journal}
  {\bibinfo  {journal} {Phys. Rev. Research}\ }\textbf {\bibinfo {volume} {3}}
  (\bibinfo {year} {2021})}\BibitemShut {NoStop}%
\bibitem [{\citenamefont {Hirzler}\ \emph {et~al.}(2022)\citenamefont
  {Hirzler}, \citenamefont {Lous}, \citenamefont {Trimby}, \citenamefont
  {P{\'{e}}rez-R{\'{\i}}os}, \citenamefont {Safavi-Naini},\ and\ \citenamefont
  {Gerritsma}}]{Hirzler2022}%
  \BibitemOpen
  \bibfield  {author} {\bibinfo {author} {\bibfnamefont {H.}~\bibnamefont
  {Hirzler}}, \bibinfo {author} {\bibfnamefont {R.}~\bibnamefont {Lous}},
  \bibinfo {author} {\bibfnamefont {E.}~\bibnamefont {Trimby}}, \bibinfo
  {author} {\bibfnamefont {J.}~\bibnamefont {P{\'{e}}rez-R{\'{\i}}os}},
  \bibinfo {author} {\bibfnamefont {A.}~\bibnamefont {Safavi-Naini}}, \ and\
  \bibinfo {author} {\bibfnamefont {R.}~\bibnamefont {Gerritsma}},\ }\href
  {\doibase 10.1103/physrevlett.128.103401} {\bibfield  {journal} {\bibinfo
  {journal} {Phys.~Rev.~Lett.}\ }\textbf {\bibinfo {volume} {128}},\ \bibinfo
  {pages} {103401} (\bibinfo {year} {2022})}\BibitemShut {NoStop}%
\bibitem [{\citenamefont {Katz}\ \emph
  {et~al.}(2022{\natexlab{a}})\citenamefont {Katz}, \citenamefont {Pinkas},
  \citenamefont {Akerman},\ and\ \citenamefont {Ozeri}}]{Katz2021}%
  \BibitemOpen
  \bibfield  {author} {\bibinfo {author} {\bibfnamefont {O.}~\bibnamefont
  {Katz}}, \bibinfo {author} {\bibfnamefont {M.}~\bibnamefont {Pinkas}},
  \bibinfo {author} {\bibfnamefont {N.}~\bibnamefont {Akerman}}, \ and\
  \bibinfo {author} {\bibfnamefont {R.}~\bibnamefont {Ozeri}},\ }\href
  {\doibase 10.1038/s41567-022-01517-y} {\bibfield  {journal} {\bibinfo
  {journal} {Nat. Phys.}\ }\textbf {\bibinfo {volume} {18}},\ \bibinfo {pages}
  {533} (\bibinfo {year} {2022}{\natexlab{a}})}\BibitemShut {NoStop}%
\bibitem [{\citenamefont {Cetina}\ \emph {et~al.}(2012)\citenamefont {Cetina},
  \citenamefont {Grier},\ and\ \citenamefont {Vuleti{\'c}}}]{Cetina:2012}%
  \BibitemOpen
  \bibfield  {author} {\bibinfo {author} {\bibfnamefont {M.}~\bibnamefont
  {Cetina}}, \bibinfo {author} {\bibfnamefont {A.~T.}\ \bibnamefont {Grier}}, \
  and\ \bibinfo {author} {\bibfnamefont {V.}~\bibnamefont {Vuleti{\'c}}},\
  }\href {\doibase 10.1103/PhysRevLett.109.253201} {\bibfield  {journal}
  {\bibinfo  {journal} {Phys. Rev. Lett.}\ }\textbf {\bibinfo {volume} {109}},\
  \bibinfo {pages} {253201} (\bibinfo {year} {2012})}\BibitemShut {NoStop}%
\bibitem [{\citenamefont {Meir}\ \emph {et~al.}(2017)\citenamefont {Meir},
  \citenamefont {Sikorsky}, \citenamefont {Ben-shlomi}, \citenamefont
  {Akerman}, \citenamefont {Pinkas}, \citenamefont {Dallal},\ and\
  \citenamefont {Ozeri}}]{meir2017}%
  \BibitemOpen
  \bibfield  {author} {\bibinfo {author} {\bibfnamefont {Z.}~\bibnamefont
  {Meir}}, \bibinfo {author} {\bibfnamefont {T.}~\bibnamefont {Sikorsky}},
  \bibinfo {author} {\bibfnamefont {R.}~\bibnamefont {Ben-shlomi}}, \bibinfo
  {author} {\bibfnamefont {N.}~\bibnamefont {Akerman}}, \bibinfo {author}
  {\bibfnamefont {M.}~\bibnamefont {Pinkas}}, \bibinfo {author} {\bibfnamefont
  {Y.}~\bibnamefont {Dallal}}, \ and\ \bibinfo {author} {\bibfnamefont
  {R.}~\bibnamefont {Ozeri}},\ }\href {\doibase 10.1080/09500340.2017.1397217}
  {\bibfield  {journal} {\bibinfo  {journal} {Journal of Modern Optics}\
  }\textbf {\bibinfo {volume} {65}},\ \bibinfo {pages} {501} (\bibinfo {year}
  {2017})}\BibitemShut {NoStop}%
\bibitem [{\citenamefont {Tomza}\ \emph {et~al.}(2019)\citenamefont {Tomza},
  \citenamefont {Jachymski}, \citenamefont {Gerritsma}, \citenamefont
  {Negretti}, \citenamefont {Calarco}, \citenamefont {Idziaszek},\ and\
  \citenamefont {Julienne}}]{Tomza:2019}%
  \BibitemOpen
  \bibfield  {author} {\bibinfo {author} {\bibfnamefont {M.}~\bibnamefont
  {Tomza}}, \bibinfo {author} {\bibfnamefont {K.}~\bibnamefont {Jachymski}},
  \bibinfo {author} {\bibfnamefont {R.}~\bibnamefont {Gerritsma}}, \bibinfo
  {author} {\bibfnamefont {A.}~\bibnamefont {Negretti}}, \bibinfo {author}
  {\bibfnamefont {T.}~\bibnamefont {Calarco}}, \bibinfo {author} {\bibfnamefont
  {Z.}~\bibnamefont {Idziaszek}}, \ and\ \bibinfo {author} {\bibfnamefont
  {P.~S.}\ \bibnamefont {Julienne}},\ }\href {\doibase
  10.1103/revmodphys.91.035001} {\bibfield  {journal} {\bibinfo  {journal}
  {Rev. Mod. Phys.}\ }\textbf {\bibinfo {volume} {91}},\ \bibinfo {pages}
  {035001} (\bibinfo {year} {2019})}\BibitemShut {NoStop}%
\bibitem [{\citenamefont {P{\'{e}}rez-R{\'{i}}os}(2020)}]{BookJPR}%
  \BibitemOpen
  \bibfield  {author} {\bibinfo {author} {\bibfnamefont {J.}~\bibnamefont
  {P{\'{e}}rez-R{\'{i}}os}},\ }\href
  {https://www.ebook.de/de/product/39345358/jesus_perez_rios_an_introduction_to_cold_and_ultracold_chemistry.html}
  {\emph {\bibinfo {title} {An Introduction to Cold and Ultracold Chemistry}}}\
  (\bibinfo  {publisher} {Springer International Publishing},\ \bibinfo {year}
  {2020})\BibitemShut {NoStop}%
\bibitem [{\citenamefont {Schneider}\ \emph {et~al.}(2010)\citenamefont
  {Schneider}, \citenamefont {Enderlein}, \citenamefont {Huber},\ and\
  \citenamefont {Schaetz}}]{schneider:2010}%
  \BibitemOpen
  \bibfield  {author} {\bibinfo {author} {\bibfnamefont {C.}~\bibnamefont
  {Schneider}}, \bibinfo {author} {\bibfnamefont {M.}~\bibnamefont
  {Enderlein}}, \bibinfo {author} {\bibfnamefont {T.}~\bibnamefont {Huber}}, \
  and\ \bibinfo {author} {\bibfnamefont {T.}~\bibnamefont {Schaetz}},\ }\href
  {https://www.nature.com/articles/nphoton.2010.236} {\bibfield  {journal}
  {\bibinfo  {journal} {Nat. Photonics}\ }\textbf {\bibinfo {volume} {4}},\
  \bibinfo {pages} {772} (\bibinfo {year} {2010})}\BibitemShut {NoStop}%
\bibitem [{\citenamefont {Leibfried}\ \emph {et~al.}(2003)\citenamefont
  {Leibfried}, \citenamefont {Blatt}, \citenamefont {Monroe},\ and\
  \citenamefont {Wineland}}]{Leibfried:2003}%
  \BibitemOpen
  \bibfield  {author} {\bibinfo {author} {\bibfnamefont {D.}~\bibnamefont
  {Leibfried}}, \bibinfo {author} {\bibfnamefont {R.}~\bibnamefont {Blatt}},
  \bibinfo {author} {\bibfnamefont {C.}~\bibnamefont {Monroe}}, \ and\ \bibinfo
  {author} {\bibfnamefont {D.}~\bibnamefont {Wineland}},\ }\href {\doibase
  10.1103/RevModPhys.75.281} {\bibfield  {journal} {\bibinfo  {journal}
  {Rev.~Mod.~Phys.}\ }\textbf {\bibinfo {volume} {75}},\ \bibinfo {pages} {281}
  (\bibinfo {year} {2003})}\BibitemShut {NoStop}%
\bibitem [{\citenamefont {F{\"u}rst}\ \emph {et~al.}(2018)\citenamefont
  {F{\"u}rst}, \citenamefont {Ewald}, \citenamefont {Secker}, \citenamefont
  {Joger}, \citenamefont {Feldker},\ and\ \citenamefont
  {Gerritsma}}]{Fuerst:2018}%
  \BibitemOpen
  \bibfield  {author} {\bibinfo {author} {\bibfnamefont {H.}~\bibnamefont
  {F{\"u}rst}}, \bibinfo {author} {\bibfnamefont {N.~V.}\ \bibnamefont
  {Ewald}}, \bibinfo {author} {\bibfnamefont {T.}~\bibnamefont {Secker}},
  \bibinfo {author} {\bibfnamefont {J.}~\bibnamefont {Joger}}, \bibinfo
  {author} {\bibfnamefont {T.}~\bibnamefont {Feldker}}, \ and\ \bibinfo
  {author} {\bibfnamefont {R.}~\bibnamefont {Gerritsma}},\ }\href
  {http://iopscience.iop.org/10.1088/1361-6455/aadd7d} {\bibfield  {journal}
  {\bibinfo  {journal} {J. Phys. B}\ }\textbf {\bibinfo {volume} {51}},\
  \bibinfo {pages} {195001} (\bibinfo {year} {2018})}\BibitemShut {NoStop}%
\bibitem [{\citenamefont {Hirzler}\ \emph {et~al.}(2020)\citenamefont
  {Hirzler}, \citenamefont {Trimby}, \citenamefont {Lous}, \citenamefont
  {Groenenboom}, \citenamefont {Gerritsma},\ and\ \citenamefont
  {P\'erez-R\'{\i}os}}]{Hirzler2020}%
  \BibitemOpen
  \bibfield  {author} {\bibinfo {author} {\bibfnamefont {H.}~\bibnamefont
  {Hirzler}}, \bibinfo {author} {\bibfnamefont {E.}~\bibnamefont {Trimby}},
  \bibinfo {author} {\bibfnamefont {R.~S.}\ \bibnamefont {Lous}}, \bibinfo
  {author} {\bibfnamefont {G.~C.}\ \bibnamefont {Groenenboom}}, \bibinfo
  {author} {\bibfnamefont {R.}~\bibnamefont {Gerritsma}}, \ and\ \bibinfo
  {author} {\bibfnamefont {J.}~\bibnamefont {P\'erez-R\'{\i}os}},\ }\href
  {\doibase 10.1103/PhysRevResearch.2.033232} {\bibfield  {journal} {\bibinfo
  {journal} {Phys. Rev. Research}\ }\textbf {\bibinfo {volume} {2}},\ \bibinfo
  {pages} {033232} (\bibinfo {year} {2020})}\BibitemShut {NoStop}%
\bibitem [{\citenamefont {Trimby}\ \emph {et~al.}(2022)\citenamefont {Trimby},
  \citenamefont {Hirzler}, \citenamefont {Fürst}, \citenamefont
  {Safavi-Naini}, \citenamefont {Gerritsma},\ and\ \citenamefont
  {Lous}}]{Trimby2022}%
  \BibitemOpen
  \bibfield  {author} {\bibinfo {author} {\bibfnamefont {E.}~\bibnamefont
  {Trimby}}, \bibinfo {author} {\bibfnamefont {H.}~\bibnamefont {Hirzler}},
  \bibinfo {author} {\bibfnamefont {H.}~\bibnamefont {Fürst}}, \bibinfo
  {author} {\bibfnamefont {A.}~\bibnamefont {Safavi-Naini}}, \bibinfo {author}
  {\bibfnamefont {R.}~\bibnamefont {Gerritsma}}, \ and\ \bibinfo {author}
  {\bibfnamefont {R.~S.}\ \bibnamefont {Lous}},\ }\href {\doibase
  10.1088/1367-2630/ac5759} {\bibfield  {journal} {\bibinfo  {journal} {New
  J.~Phys.}\ }\textbf {\bibinfo {volume} {24}},\ \bibinfo {pages} {035004}
  (\bibinfo {year} {2022})}\BibitemShut {NoStop}%
\bibitem [{Note1()}]{Note1}%
  \BibitemOpen
  \bibinfo {note} {At the inner turning point, the forces maximize and the
  step-size reaches its lowest values, which reaches below an identifier
  value.}\BibitemShut {Stop}%
\bibitem [{\citenamefont {Weckesser}\ \emph {et~al.}(2021)\citenamefont
  {Weckesser}, \citenamefont {Thielemann}, \citenamefont {Wiater},
  \citenamefont {Wojciechowska}, \citenamefont {Karpa}, \citenamefont
  {Jachymski}, \citenamefont {Tomza}, \citenamefont {Walker},\ and\
  \citenamefont {Schaetz}}]{Weckesser:2021}%
  \BibitemOpen
  \bibfield  {author} {\bibinfo {author} {\bibfnamefont {P.}~\bibnamefont
  {Weckesser}}, \bibinfo {author} {\bibfnamefont {F.}~\bibnamefont
  {Thielemann}}, \bibinfo {author} {\bibfnamefont {D.}~\bibnamefont {Wiater}},
  \bibinfo {author} {\bibfnamefont {A.}~\bibnamefont {Wojciechowska}}, \bibinfo
  {author} {\bibfnamefont {L.}~\bibnamefont {Karpa}}, \bibinfo {author}
  {\bibfnamefont {K.}~\bibnamefont {Jachymski}}, \bibinfo {author}
  {\bibfnamefont {M.}~\bibnamefont {Tomza}}, \bibinfo {author} {\bibfnamefont
  {T.}~\bibnamefont {Walker}}, \ and\ \bibinfo {author} {\bibfnamefont
  {T.}~\bibnamefont {Schaetz}},\ }\href {\doibase 10.1038/s41586-021-04112-y}
  {\bibfield  {journal} {\bibinfo  {journal} {Nature}\ }\textbf {\bibinfo
  {volume} {600}},\ \bibinfo {pages} {429} (\bibinfo {year}
  {2021})}\BibitemShut {NoStop}%
\bibitem [{\citenamefont {Hippler}\ \emph {et~al.}(1990)\citenamefont
  {Hippler}, \citenamefont {Rahn},\ and\ \citenamefont {Troe}}]{ozone1}%
  \BibitemOpen
  \bibfield  {author} {\bibinfo {author} {\bibfnamefont {H.}~\bibnamefont
  {Hippler}}, \bibinfo {author} {\bibfnamefont {R.}~\bibnamefont {Rahn}}, \
  and\ \bibinfo {author} {\bibfnamefont {J.}~\bibnamefont {Troe}},\ }\href
  {\doibase 10.1063/1.458972} {\bibfield  {journal} {\bibinfo  {journal} {The
  Journal of Chemical Physics}\ }\textbf {\bibinfo {volume} {93}},\ \bibinfo
  {pages} {6560} (\bibinfo {year} {1990})},\ \Eprint
  {http://arxiv.org/abs/https://doi.org/10.1063/1.458972}
  {https://doi.org/10.1063/1.458972} \BibitemShut {NoStop}%
\bibitem [{\citenamefont {Mirahmadi}\ \emph {et~al.}(2022)\citenamefont
  {Mirahmadi}, \citenamefont {P\'erez-R\'{\i}os}, \citenamefont {Egorov},
  \citenamefont {Tyuterev},\ and\ \citenamefont {Kokoouline}}]{ozone2}%
  \BibitemOpen
  \bibfield  {author} {\bibinfo {author} {\bibfnamefont {M.}~\bibnamefont
  {Mirahmadi}}, \bibinfo {author} {\bibfnamefont {J.}~\bibnamefont
  {P\'erez-R\'{\i}os}}, \bibinfo {author} {\bibfnamefont {O.}~\bibnamefont
  {Egorov}}, \bibinfo {author} {\bibfnamefont {V.}~\bibnamefont {Tyuterev}}, \
  and\ \bibinfo {author} {\bibfnamefont {V.}~\bibnamefont {Kokoouline}},\
  }\href {\doibase 10.1103/PhysRevLett.128.108501} {\bibfield  {journal}
  {\bibinfo  {journal} {Phys. Rev. Lett.}\ }\textbf {\bibinfo {volume} {128}},\
  \bibinfo {pages} {108501} (\bibinfo {year} {2022})}\BibitemShut {NoStop}%
\bibitem [{\citenamefont {Roberts}\ \emph {et~al.}(1969)\citenamefont
  {Roberts}, \citenamefont {Bernstein},\ and\ \citenamefont
  {Curtiss}}]{RBCmechanism}%
  \BibitemOpen
  \bibfield  {author} {\bibinfo {author} {\bibfnamefont {R.~E.}\ \bibnamefont
  {Roberts}}, \bibinfo {author} {\bibfnamefont {R.~B.}\ \bibnamefont
  {Bernstein}}, \ and\ \bibinfo {author} {\bibfnamefont {C.~F.}\ \bibnamefont
  {Curtiss}},\ }\href {\doibase 10.1063/1.1671032} {\bibfield  {journal}
  {\bibinfo  {journal} {The Journal of Chemical Physics}\ }\textbf {\bibinfo
  {volume} {50}},\ \bibinfo {pages} {5163} (\bibinfo {year} {1969})},\ \Eprint
  {http://arxiv.org/abs/https://doi.org/10.1063/1.1671032}
  {https://doi.org/10.1063/1.1671032} \BibitemShut {NoStop}%
\bibitem [{\citenamefont {Orel}(1987)}]{RBCmechanism2}%
  \BibitemOpen
  \bibfield  {author} {\bibinfo {author} {\bibfnamefont {A.~E.}\ \bibnamefont
  {Orel}},\ }\href {\doibase 10.1063/1.453628} {\bibfield  {journal} {\bibinfo
  {journal} {The Journal of Chemical Physics}\ }\textbf {\bibinfo {volume}
  {87}},\ \bibinfo {pages} {314} (\bibinfo {year} {1987})},\ \Eprint
  {http://arxiv.org/abs/https://doi.org/10.1063/1.453628}
  {https://doi.org/10.1063/1.453628} \BibitemShut {NoStop}%
\bibitem [{\citenamefont {P{\'{e}}rez-R{\'{\i}}os}\ and\ \citenamefont
  {Greene}(2015)}]{perezrios2015}%
  \BibitemOpen
  \bibfield  {author} {\bibinfo {author} {\bibfnamefont {J.}~\bibnamefont
  {P{\'{e}}rez-R{\'{\i}}os}}\ and\ \bibinfo {author} {\bibfnamefont {C.~H.}\
  \bibnamefont {Greene}},\ }\href {\doibase 10.1063/1.4927702} {\bibfield
  {journal} {\bibinfo  {journal} {J. Chem. Phys.}\ }\textbf {\bibinfo {volume}
  {143}},\ \bibinfo {pages} {041105} (\bibinfo {year} {2015})}\BibitemShut
  {NoStop}%
\bibitem [{\citenamefont {Krükow}\ \emph {et~al.}(2016)\citenamefont
  {Krükow}, \citenamefont {Mohammadi}, \citenamefont {Härter},\ and\
  \citenamefont {Denschlag}}]{kruekow2016}%
  \BibitemOpen
  \bibfield  {author} {\bibinfo {author} {\bibfnamefont {A.}~\bibnamefont
  {Krükow}}, \bibinfo {author} {\bibfnamefont {A.}~\bibnamefont {Mohammadi}},
  \bibinfo {author} {\bibfnamefont {A.}~\bibnamefont {Härter}}, \ and\
  \bibinfo {author} {\bibfnamefont {J.~H.}\ \bibnamefont {Denschlag}},\ }\href
  {https://journals.aps.org/pra/abstract/10.1103/PhysRevA.94.030701} {\bibfield
   {journal} {\bibinfo  {journal} {Phys.~Rev.~A}\ }\textbf {\bibinfo {volume}
  {94}} (\bibinfo {year} {2016})}\BibitemShut {NoStop}%
\bibitem [{\citenamefont {P{\'{e}}rez-R{\'{\i}}os}\ and\ \citenamefont
  {Greene}(2018)}]{perezrios2018}%
  \BibitemOpen
  \bibfield  {author} {\bibinfo {author} {\bibfnamefont {J.}~\bibnamefont
  {P{\'{e}}rez-R{\'{\i}}os}}\ and\ \bibinfo {author} {\bibfnamefont {C.~H.}\
  \bibnamefont {Greene}},\ }\href {\doibase 10.1103/physreva.98.062707}
  {\bibfield  {journal} {\bibinfo  {journal} {Phys. Rev. A}\ }\textbf {\bibinfo
  {volume} {98}},\ \bibinfo {pages} {062707} (\bibinfo {year}
  {2018})}\BibitemShut {NoStop}%
\bibitem [{\citenamefont {Katz}\ \emph
  {et~al.}(2022{\natexlab{b}})\citenamefont {Katz}, \citenamefont {Pinkas},
  \citenamefont {Akerman},\ and\ \citenamefont {Ozeri}}]{Katz2022}%
  \BibitemOpen
  \bibfield  {author} {\bibinfo {author} {\bibfnamefont {O.}~\bibnamefont
  {Katz}}, \bibinfo {author} {\bibfnamefont {M.}~\bibnamefont {Pinkas}},
  \bibinfo {author} {\bibfnamefont {N.}~\bibnamefont {Akerman}}, \ and\
  \bibinfo {author} {\bibfnamefont {R.}~\bibnamefont {Ozeri}},\ }\href@noop {}
  {\  (\bibinfo {year} {2022}{\natexlab{b}})},\ \Eprint
  {http://arxiv.org/abs/2208.07725} {arXiv:2208.07725 [quant-ph]} \BibitemShut
  {NoStop}%
\bibitem [{\citenamefont {Pinkas}\ \emph {et~al.}(2022)\citenamefont {Pinkas},
  \citenamefont {Katz}, \citenamefont {Wengrowicz}, \citenamefont {Akerman},\
  and\ \citenamefont {Ozeri}}]{Pinkas2022}%
  \BibitemOpen
  \bibfield  {author} {\bibinfo {author} {\bibfnamefont {M.}~\bibnamefont
  {Pinkas}}, \bibinfo {author} {\bibfnamefont {O.}~\bibnamefont {Katz}},
  \bibinfo {author} {\bibfnamefont {J.}~\bibnamefont {Wengrowicz}}, \bibinfo
  {author} {\bibfnamefont {N.}~\bibnamefont {Akerman}}, \ and\ \bibinfo
  {author} {\bibfnamefont {R.}~\bibnamefont {Ozeri}},\ }\href@noop {} {\
  (\bibinfo {year} {2022})},\ \Eprint {http://arxiv.org/abs/2208.06904}
  {arXiv:2208.06904 [physics.atom-ph]} \BibitemShut {NoStop}%
\end{thebibliography}

%

\end{document}